\documentclass[11pt]{article}
\usepackage[a4paper,margin=27mm]{geometry}
\usepackage{amsmath,amssymb,amsthm}
\usepackage{booktabs}
\usepackage[hidelinks]{hyperref}
\newtheorem{theorem}{Theorem}
\newtheorem{lemma}{Lemma}
\newtheorem{proposition}{Proposition}
\newcommand{\KL}{D}
\newcommand{\E}{\mathbb E}
\newcommand{\R}{\mathcal R}
\newcommand{\ind}{\mathbf 1}
\newcommand{\N}{\mathbb N}
\title{Polynomial-Time Minimax Prediction for Stationary Renewal Processes\thanks{OpenAI Codex assisted with mathematical derivations, software development, computational checks, and manuscript preparation. The authors are responsible for the final content.}}
\author{Xinyu Chen \and Yiming Ma}
\date{1 November 2026}
\hypersetup{pdftitle={Polynomial-Time Minimax Prediction for Stationary Renewal Processes},
pdfauthor={Xinyu Chen and Yiming Ma}}
\begin{document}
\maketitle
\begin{abstract}
The minimax next-symbol Kullback--Leibler risk for stationary binary
renewal processes is known to have order $n^{-1/2}$. We give a
polynomial-time predictor attaining this order for an arbitrary unknown
finite-mean interarrival law. The construction uses a classical
type-uniform renewal code. Its contribution is an exact dynamic program
for censored prefix marginals, avoiding enumeration of gap histograms
or binary completions. Combined with an established prediction reduction,
it gives risk at most $[\log M_{n+1}+\log(n+2)]/n$, where $M_N$ counts
renewal types. Shared suffix updates reduce the total computation to
$O(n^3)$ arithmetic work and polynomial
bit complexity, without tail, mixing, known-mean, or bounded-support
assumptions. The statistical rate and its lower bound are prior results;
the proposed advance concerns computation.
\end{abstract}

\section{Problem and contribution}
Let $T_1,T_2,\ldots$ be independent with common law $\mu$ on
$\N=\{1,2,\ldots\}$ and finite mean $m_\mu$. Independently choose $T_0$
with the equilibrium residual law
\[
 \Pr(T_0=a)=R_\mu(a)/m_\mu,\qquad R_\mu(a)=\Pr_\mu(T_1\ge a).
\]
Set $X_i=1$ exactly at times $T_0,T_0+T_1,\ldots$. These laws form the
stationary binary renewal class $\R$. Neither $\mu$ nor its mean is known.
Given $X_1^n=(X_1,\ldots,X_n)$, define next-symbol prediction risk by
\[
 {\cal L}_n(P,q_n)=
 \E_P\KL\bigl(P(X_{n+1}\in\cdot\mid X_1^n)
                  \,\|\,q_n(\cdot\mid X_1^n)\bigr).
\]
Logarithms are natural. Han, Jiang, and Wu \cite{hjw24} proved minimax
risk of order $n^{-1/2}$ and explicitly asked for a polynomial-time
predictor attaining this order.

The type method and $O(\sqrt n)$ renewal coding bound are classical
\cite{cs96,cs98,fs02}. Computing a complete-word probability, however,
does not compute its prefix marginals. We give the latter computation
for the classical type-uniform code and apply the suffix-averaged
prediction reduction of \cite{hjw24}. No new statistical minimax rate,
lower bound, sufficient statistic, or general reduction is claimed.

The computation must account for a last observed gap that has not yet
ended. Direct enumeration can visit superpolynomially many gap histograms;
we instead retain coefficients indexed only by future span and number
of gaps. A marked coefficient records whether the first future gap
exceeds the observed age. Removing observations from the left then
updates these coefficients by multiplication by a single polynomial,
allowing all suffixes to share one computation.

Let $p(s)$ be the integer partition number, $p(0)=1$, and set
\begin{equation}
 M_0=1,\qquad M_N=1+\sum_{s=0}^{N-1}(N-s)p(s)\quad(N\ge1).
 \label{eq:types}
\end{equation}
\begin{theorem}\label{thm:main}
There is an explicit deterministic rational-valued predictor $q_n$
such that for every $n\ge1$,
\[
 \sup_{P\in\R}{\cal L}_n(P,q_n)
 \le\frac{\log M_{n+1}+\log(n+2)}n=O(n^{-1/2}).
\]
Its rate is minimax optimal by \cite[Theorem 1.4]{hjw24}.
The predictor is evaluated by the algorithm in Theorem~\ref{thm:algorithm}.
\end{theorem}
The statistical consequence and the new computational statement are
separate: Theorem~\ref{thm:algorithm} evaluates all required prefix masses
in $O(n^3)$ arithmetic operations, with polynomial bit complexity and
$O(n^2)$ stored integer entries. Theorem~\ref{thm:main} then applies an
existing statistical reduction to that evaluator.
Polynomial complexity is not a large-scale runtime claim. The predictor
may use the observation horizon; no horizon-free update result is asserted.

This work resolves the computational question raised by Han, Jiang,
and Wu (2024)~\cite{hjw24} concerning the existence of a polynomial-time
predictor attaining the minimax-optimal risk for stationary renewal processes.

\section{Classical renewal types}
For a nonzero word $x\in\{0,1\}^N$, let $a,b$ be its first and last one
positions, and $c_j$ the number of completed gaps of length $j$.
Put $m=\sum_jc_j$. Its type is $(a,b,c)$; the all-zero word is a
separate type. Ordering the completed gaps proves
\[
 \sum_jjc_j=b-a,\qquad |{\cal T}(a,b,c)|=\frac{m!}{\prod_jc_j!}.
\]
Each ordering gives a unique word. There are $p(s)$ histograms of span
$s$ and $N-s$ choices for $a$, proving \eqref{eq:types}.
The classical type-uniform distribution is
\begin{equation}
 Q_N(x)=\frac1{M_N|{\cal T}(x)|}.                       \label{eq:code}
\end{equation}
\begin{lemma}[Classical coding bound]\label{lem:code}
For every $P\in\R$,
\[
 P(x)\le M_NQ_N(x),\qquad
 \KL(P_{X_1^N}\|Q_N)\le\log M_N
 \le\pi\sqrt{2N/3}+2\log(N+1).
\]
\end{lemma}
\begin{proof}
Every nonzero word has probability
\[
 P(x)=\frac{R_\mu(a)}{m_\mu}
           \prod_j\mu(j)^{c_j}R_\mu(N-b+1).
\]
It is constant on a type, hence $P(x)|{\cal T}(x)|\le1$. The zero
type is a singleton. This proves the first two inequalities.
For $\theta>0$, Euler's partition product gives
\[
 \log\sum_{s\ge0}p(s)e^{-\theta s}
 =\sum_{k\ge1}\frac1{k(e^{\theta k}-1)}
 \le\frac{\pi^2}{6\theta}.
\]
Optimizing $\theta s+\pi^2/(6\theta)$ bounds $p(s)$ by
$\exp(\pi\sqrt{2s/3})$ for $s>0$. Equation \eqref{eq:types} then
gives the final bound.
\end{proof}
The type bound also holds for arbitrary independent initial waiting
times; stationarity will enter the prediction argument separately.

\section{Censored prefix marginals}
This section supplies the computational result. We first express a
prefix mass as a sum over admissible future gap counts, then evaluate
its coefficients, and finally share these evaluations across suffixes.
The observation horizon stays fixed in all three steps.
Fix one horizon $N$. For a prefix $u$ of length $t\le N$, define
\[
 W_N(u)=\sum_{x_1^t=u}\frac1{|{\cal T}(x)|},\qquad S_N(u)=N!W_N(u).
\]
Thus $Q_N(u)=S_N(u)/(N!M_N)$. When $u=0^t$, compatible types are
exactly the length-$(N-t)$ types translated by $t$, including the zero
type. Each contributes one, so
\begin{equation}
 W_N(0^t)=M_{N-t}.                                    \label{eq:zeros}
\end{equation}
Otherwise let $a,b,c,m$ describe the observed completed gaps and set
$d=t-b$, $R=N-b$, and $C=\prod_jc_j!$. The completion with no further
one has weight $C/m!$. For future gap counts $e_j$, write
$k=\sum_je_j$ and $s=\sum_jje_j\le R$. The first future gap must
exceed the censored age $d$. Define
\begin{equation}
 B_{s,k}=\sum_{\substack{\sum j e_j=s\\\sum e_j=k}}
       \left(\sum_{j>d}e_j\right)\prod_j\binom{c_j+e_j}{e_j}.
 \label{eq:B}
\end{equation}
\begin{lemma}[Prefix identity]\label{lem:prefix}
\begin{align}
 W_N(u)&=\frac C{m!}
       +C\sum_{s=1}^R\sum_{k=1}^s\frac{(k-1)!}{(m+k)!}B_{s,k},
                                                        \label{eq:W}\\
 S_N(u)&=\frac{CN!}{m!}
       +C\sum_{s=1}^R\sum_{k=1}^s
                 B_{s,k}(k-1)!\frac{N!}{(m+k)!}.         \label{eq:S}
\end{align}
All terms in \eqref{eq:S} are nonnegative integers.
\end{lemma}
\begin{proof}
For a fixed histogram $e$, choose its first gap $j>d$. The number of
allowed orderings is
\[
 \sum_{\substack{j>d\\e_j>0}}
 \frac{(k-1)!}{(e_j-1)!\prod_{i\ne j}e_i!}
 =\frac{(k-1)!}{\prod_je_j!}\sum_{j>d}e_j .
\]
Every ordering determines the future one positions and then the trailing
zeros. The full word has inverse type size
$\prod_j(c_j+e_j)!/(m+k)!$. Multiplication and summation give
\eqref{eq:W}. Also $m+k\le N-1$, since the first one is at $a\ge1$,
proving integrality in \eqref{eq:S}. Counts with $j>R$ remain in
$C,m$, although their future multiplicities are zero.
\end{proof}

The strict inequality $j>d$ in the mark is essential. A gap of length
$d$ would put the next one at the last observed position, contradicting
the observed trailing zeros. The zero-future-gap term $C/m!$ must also
be retained: the observation window may end before the next renewal.
For a single observed one, $m=0$ and $C=1$; the same identity applies
without an artificial completed gap before the first observation.
Repeated gaps enter through their multiplicities $c_j$, including those
with $j>R$. Such counts affect the full type's factorial weight even
though a future gap of that length cannot fit in the remaining window.

\subsection{Two coefficient recurrences}
Initialize arrays $A,B$ on $0\le k\le s\le R$ with $A_{0,0}=1$
and other entries zero. Process gap lengths $j=1,\ldots,R$. For
each old state $(s,k)$ and $0\le e\le\lfloor(R-s)/j\rfloor$,
set $g=\binom{c_j+e}{e}$ and update new arrays by
\begin{align*}
 A'_{s+je,k+e}&\mathrel{+}=gA_{s,k},\\
 B'_{s+je,k+e}&\mathrel{+}=g(B_{s,k}+e\ind_{\{j>d\}}A_{s,k}).
\end{align*}
Induction on $j$ shows that $A$ sums products of binomial factors
and $B$ additionally counts the gaps longer than $d$. Thus the final
$B$ is exactly \eqref{eq:B}. Equivalently, it is the coefficient
array of
\[
 G(z,w)\sum_{j=d+1}^R\frac{(c_j+1)wz^j}{1-wz^j},
 \qquad G(z,w)=\prod_{j=1}^R(1-wz^j)^{-c_j-1}.
\]

The explicit multiplicity loop costs $O(N^4)$ arithmetic operations
per prefix. To reduce this to $O(N^3)$, replace each factor by $c_j+1$
copies of $(1-wz^j)^{-1}$. For each copy visit $s$ in increasing
order and, for $k\ge1$, update in place:
\begin{align*}
 A_{s,k}&\leftarrow A_{s,k}+A_{s-j,k-1},\\
 B_{s,k}&\leftarrow B_{s,k}+B_{s-j,k-1}
                            +\ind_{\{j>d\}}A_{s-j,k-1}.
\end{align*}
The entries at $s-j$ already include the current factor. These are
its ordinary and marked coefficient recurrences. The number of
factors is at most $R+m\le N-a\le N-1$, with $O(N^2)$ states.

\subsection{Sharing the suffix evaluations}
\begin{theorem}[Simultaneous fixed-horizon prefix evaluation]
\label{thm:algorithm}
For every binary history $x_1^n$, $n\ge1$, put $N=n+1$ and
$u_t=x_{n-t+1}^n$. All $2n$ positive integers
\begin{equation}
 D_t=S_N(u_t),\qquad E_t=S_N(u_t1),\qquad 1\le t\le n,
 \label{eq:DE}
\end{equation}
can be computed exactly in a total of $O(n^3)$ arithmetic operations.
The algorithm stores $O(n^2)$ integer entries. Its coefficient entries
have $O(n)$ bits, and its prefix masses have $O(n\log(n+1))$ bits.
Consequently all conditionals
\begin{equation}
 \rho_t=Q_N(1\mid u_t)=E_t/D_t
 \label{eq:fixedconditional}
\end{equation}
and their average can be computed in polynomial bit time. The average
requires $O(n)$ further arithmetic operations if fractions are added
without cancellation.
\end{theorem}
The proof is constructive. The coefficient initialization, queries, and
deletions below constitute Algorithms~1--3; their invariants establish
\eqref{eq:DE}, and the subsequent bit analysis completes the theorem.

Let $d$ be the number of trailing zeros in the length-$n$ history.
Write $u_t=x_{n-t+1}^n$ for its suffix of length $t$.
If $d=n$, every suffix conditional follows from \eqref{eq:zeros}.
Otherwise all nonzero suffixes $u_t$, $t>d$, have the same age $d$.
For each such suffix keep its histogram $c$, gap count $m$, and
$C=\prod_j c_j!$. Put $h=d+1$. Appending a one adds precisely one
gap of length $h$, so $m^+=m+1$ and $C^+=C(c_h+1)$.

To reuse computations as suffixes shorten, retain all coefficients
through span $N$ in the three formal series
\begin{align*}
 \widetilde G_c&=\prod_{j=1}^N(1-wz^j)^{-c_j-1},\\
 \widetilde H_{c,d}
  &=\widetilde G_c\sum_{j=d+1}^N\frac{(c_j+1)wz^j}{1-wz^j},
 &\widetilde G_c^+&=\frac{\widetilde G_c}{1-wz^h}.
\end{align*}
Denote their coefficients by $A_{s,k},B_{s,k},A^+_{s,k}$.
For $u_t$, use \eqref{eq:S} with $R=N-t+d$.
For $u_t1$, the age is zero, so its marked coefficient is $kA^+_{s,k}$,
and
\begin{equation}
 S_N(u_t1)=\frac{C^+N!}{(m+1)!}
 +C^+\sum_{s=1}^{N-t-1}\sum_{k=1}^s
                  A^+_{s,k}k!\frac{N!}{(m+1+k)!}.
 \label{eq:append}
\end{equation}
Factors of span exceeding a query budget have constant term one
and cannot affect that query. Retaining them is necessary because
the budget grows when the suffix shortens.

Deleting a leading zero leaves the arrays unchanged. Deleting a
leading one, while another one remains, removes its completed gap $j$.
Write $v=wz^j$ and $b=\ind_{\{j>d\}}$. The updated histogram $c'$ obeys
\[
 \widetilde G_{c'}=(1-v)\widetilde G_c,\qquad
 \widetilde G_{c'}^+=(1-v)\widetilde G_c^+,\qquad
 \widetilde H_{c',d}=(1-v)\widetilde H_{c,d}-bv\widetilde G_c .
\]
The last identity is the product rule for the marked series.
Thus, with all right sides referring to old coefficients,
\begin{equation}
\begin{aligned}
 A'_{s,k}&=A_{s,k}-A_{s-j,k-1},\\
 B'_{s,k}&=B_{s,k}-B_{s-j,k-1}-bA_{s-j,k-1},\\
 (A^+)'_{s,k}&=A^+_{s,k}-A^+_{s-j,k-1}.
\end{aligned}
\label{eq:delete}
\end{equation}
Descending span permits in-place updates. The remaining product
still has positive integer exponents, so its coefficients are
nonnegative. Update $C'=C/c_j$ using the old positive count and
$m'=m-1$. After the last one is removed, use \eqref{eq:zeros}:
the conditional at suffix $0^t$ is $1-M_{N-t-1}/M_{N-t}$.

Initialization uses at most $N+m+1\le2N$ geometric factors and
$O(N^2)$ states, costing $O(N^3)$ operations. Each of the $O(N)$
deletions and queries costs $O(N^2)$. This proves the total
$O(N^3)$ arithmetic bound with three $O(N^2)$ arrays.
No coefficient of span greater than $N$ can affect a retained
coefficient in any of these multiplications.

\paragraph{Algorithm 1: initialize the coefficient arrays.}
Input the horizon $N$, observed histogram $c$, and trailing age $d$.
Every array is indexed by $0\le k\le s\le N$; invalid indices mean zero.
\begin{enumerate}
\item Set all $A,B$ entries to zero and $A_{0,0}\leftarrow1$.
\item \textbf{For} $j=1,\ldots,N$, put $b\leftarrow\ind_{\{j>d\}}$.
 Repeat $c_j+1$ times: \textbf{for} $s=j,j+1,\ldots,N$ and
 $k=1,\ldots,s$, put $a\leftarrow A_{s-j,k-1}$ and
 $b'\leftarrow B_{s-j,k-1}$, then set
 \[
  A_{s,k}\leftarrow A_{s,k}+a,\qquad
  B_{s,k}\leftarrow B_{s,k}+b'+ba.
 \]
\item Copy $A^+\leftarrow A$ and put $h\leftarrow d+1$.
 \textbf{For} $s=h,h+1,\ldots,N$ and $k=1,\ldots,s$, set
 \[
 A^+_{s,k}\leftarrow A^+_{s,k}+A^+_{s-h,k-1}.
 \]
\item Return $A,B,A^+$. They represent respectively
 $\widetilde G_c,\widetilde H_{c,d},\widetilde G_c^+$ through span $N$.
\end{enumerate}
Ascending $s$ allows arbitrarily many copies of the current geometric
factor. Each cell requires a bounded number of integer operations;
there are at most $2N$ factors and $O(N^2)$ cells.

\paragraph{Algorithm 2: delete one completed gap.}
Input $j$ with $c_j>0$ and the three arrays for the old histogram.
\begin{enumerate}
\item Save $b\leftarrow\ind_{\{j>d\}}$ and $c_{\rm old}\leftarrow c_j$.
\item \textbf{For} $s=N,N-1,\ldots,j$ and $k=1,\ldots,s$,
 save $a\leftarrow A_{s-j,k-1}$ and $b'\leftarrow B_{s-j,k-1}$.
 Apply
 \[
 A_{s,k}\leftarrow A_{s,k}-a,\quad
 B_{s,k}\leftarrow B_{s,k}-b'-ba,\quad
 A^+_{s,k}\leftarrow A^+_{s,k}-A^+_{s-j,k-1}.
 \]
\item Set $C\leftarrow C/c_{\rm old}$, $c_j\leftarrow c_j-1$, and
 $m\leftarrow m-1$. Return the updated arrays and counts.
\end{enumerate}
Here descending $s$ preserves the old lower-span coefficients. The
marked correction $-ba$ is needed in addition to multiplying $B$ by
$1-wz^j$. Omitting it changes the meaning of the admissible-first-gap mark.

\paragraph{Algorithm 3: compute every prefix pair and the predictor.}
Input $x_1^n$. Put $N=n+1$, and let $F_i=i!$ for $0\le i\le N$.
\begin{enumerate}
\item Compute the partition numbers by $p_0\leftarrow1$, other $p_s=0$;
 \textbf{for} $j=1,\ldots,N$ and ascending $s=j,\ldots,N$, set
 $p_s\leftarrow p_s+p_{s-j}$. Compute the $M_i$ from \eqref{eq:types}
 and the factorials $F_i=iF_{i-1}$. This takes $O(N^2)$ operations.
\item Find the trailing age $d$. \textbf{For} $t=1,\ldots,d$, set
 \[
 D_t=F_NM_{N-t},\qquad E_t=F_N(M_{N-t}-M_{N-t-1}).
 \]
 If $d=n$, go to step~5.
\item From the ordered one positions form $c,m,C$. Run Algorithm~1.
 Keep a pointer to the first remaining one position.
\item \textbf{For} $t=n,n-1,\ldots,d+1$, set
 \[
 \begin{split}
 D_t&\leftarrow CF_N/F_m,\qquad C^+\leftarrow C(c_{d+1}+1),\\
 E_t&\leftarrow C^+F_N/F_{m+1}.
 \end{split}
 \]
 Add to $D_t$ every term
 $CB_{s,k}F_{k-1}(F_N/F_{m+k})$ for
 $1\le k\le s\le N-t+d$. Add to $E_t$ every term
 $C^+A^+_{s,k}F_k(F_N/F_{m+1+k})$ for
 $1\le k\le s\le N-t-1$.
 If $t>d+1$ and $x_{n-t+1}=1$, let $j$ be the distance to the next
 remaining one, run Algorithm~2, and advance the pointer. Deleting
 a zero changes only $t$ and hence the query budgets.
\item Return $(D_t,E_t)_{t=1}^n$ and
 $q_n(1)=n^{-1}\sum_{t=1}^nE_t/D_t$, $q_n(0)=1-q_n(1)$.
\end{enumerate}
All factorial divisions are exact. Zero coefficients may be skipped
before multiplying by a factorial ratio; they correspond to no admissible
completion. The largest denominator index is at most $N$ in every query.
The invariant at entry to step~4 is that the arrays and $c,m,C$ belong to
$u_t$, with the original horizon $N$ and the original age $d$ retained.
The prefix and append identities prove each query. Deletion preserves the
invariant by \eqref{eq:delete}; the precomputed zero branch covers the
remaining suffixes. Thus Algorithm~3 proves \eqref{eq:DE}.

\subsection{Bit complexity}
Dropping the span constraint bounds a degree-$k$ coefficient of any
ordinary series above, whose total exponent is at most $2N$,
by $ \binom{2N+k-1}{k}$. For $k\le N$, it has $O(N)$ bits; marking
adds a factor at most $N$. Moreover
\[
 0<S_N(u)\le N!M_N\le N!2^N .
\]
Deletion differences also have $O(N)$ bits. All terms in \eqref{eq:S}
and the append-one formula are nonnegative, so their products and partial
sums have $O(N\log(N+1))$ bits. Factorials and type counts have
polynomial preprocessing cost.
Rebuilding each prefix separately would cost $O(N^4)$; the shared
computation above costs $O(N^3)$. The sum of
$N-1$ rational conditionals has a denominator of at most
$O(N^2\log(N+1))$ bits. Exact integer arithmetic thus has polynomial
bit cost; no unit-cost unbounded register or exact-real oracle is assumed.
For the arithmetic bound one may add the fractions without cancellation
using $O(N)$ further operations; the same denominator bound applies.
These bounds complete the proof of Theorem~\ref{thm:algorithm}.

\section{Risk of the predictor}
Set $N=n+1$ and $u_t=x_{n-t+1}^n$. Define
\begin{equation}
 q_n(1\mid x_1^n)=\frac1n\sum_{t=1}^n
                 \frac{S_N(u_t1)}{S_N(u_t)},\qquad
 q_n(0\mid x_1^n)=1-q_n(1\mid x_1^n).                 \label{eq:q}
\end{equation}
Full support and $S_N(u)=S_N(u0)+S_N(u1)$ make every conditional
valid. All suffixes use the same horizon-$N$ distribution.
\begin{proof}[Proof of Theorem \ref{thm:main}]
The reduction of \cite[Proposition 2.1]{hjw24} gives
\begin{align}
 {\cal L}_n(P,q_n)
 &\le \frac{\KL(P_{X_1^N}\|Q_N)-\KL(P_{X_1}\|Q_N(X_1))}{n}
       \notag\\
 &\quad+\frac1n\sum_{t=1}^n
           I(X_N;X_1^{n-t}\mid X_{n-t+1}^n).          \label{eq:reduction}
\end{align}
To see this, use KL convexity in \eqref{eq:q} and insert
$P(X_N\mid X_{n-t+1}^n)$ into each likelihood ratio. Stationarity
identifies each conditional divergence with the first $t+1$
coordinates; the chain rule telescopes their sum.

As in \cite[Section 3.4]{hjw24}, represent the process by a stationary
countdown chain $Z_i$: decrement if $Z_i>1$, draw a fresh $\mu$ value
after state $1$, and emit $X_i=\ind_{\{Z_i=1\}}$. Conditional data
processing and stationarity give
\[
 I(X_N;X_1^{n-t}\mid X_{n-t+1}^n)
 \le I(X_N;Z_{n-t+1}\mid X_{n-t+1}^n)
 = I(X_{t+1};Z_1\mid X_1^t).
\]
Summation bounds the memory term's numerator by $I(Z_1;X_1^N)$.
Also $Z_1\to\min(Z_1,N+1)\to X_1^N$ is a Markov chain: all larger
countdowns generate the same $N$ zeros. Therefore
\[
 I(Z_1;X_1^N)\le H(\min(Z_1,N+1))\le\log(N+1).
\]
No bound on the entropy of $Z_1$ itself is needed. Combine this with
\eqref{eq:reduction} and Lemma \ref{lem:code}. Complexity was proved above.
\end{proof}

\section{Failure of terminal-score shortcuts}
Efficient full-word evaluation need not give an efficient generic
offline-to-online conversion \cite{hutter14}. Even within the present
model the following simple shortcuts fail.
\begin{proposition}
Normalizing only $Q_{n+1}(x_1^n0),Q_{n+1}(x_1^n1)$, or using the
terminal conditional of $Q_{n+1}$ without suffix averaging, does not
give uniformly vanishing next-symbol risk on $\R$.
\end{proposition}
\begin{proof}
For the stationary periodic source $\mu=\delta_L$, $L>n+1$, history
$0^n$ has mass $(L-n)/L$ and oracle next-one probability $1/(L-n)$.
Both words $0^{n+1}$ and $0^n1$ have singleton types. Both shortcuts
predict $1/2$ on this history; they are algebraically the same estimator.
Averaging local-horizon normalized full-word scores over suffixes also
gives $1/2$ on this history. Their risk is at least
\[
 \frac{L-n}{L}\left[\log2-h\left(\frac1{L-n}\right)\right],
\]
where $h$ is binary entropy in nats. This tends to $\log2$ for
$L=n^3$. In contrast, \eqref{eq:zeros} and marginal consistency imply
\[
 q_n(1\mid0^n)=\frac1n\sum_{r=1}^n
                  \left(1-\frac{M_{r-1}}{M_r}\right)
\]
for the actual averaged rule. Neither full-word scores nor a terminal
conditional replaces the prefix calculation and the suffix average.
\end{proof}

\section{Prior work and scope}
\subsection{Which probability assignment must be evaluated?}
The classical code is not proposed as new. For this code, the
conditional required by the reduction is exactly
\begin{equation}
 Q_N(1\mid u)=
 \frac{\displaystyle\sum_{v\in\{0,1\}^{N-|u|-1}}
                         |\mathcal T(u1v)|^{-1}}
      {\displaystyle\sum_{v\in\{0,1\}^{N-|u|}}
                         |\mathcal T(uv)|^{-1}}.
 \label{eq:requiredquery}
\end{equation}
Its horizon is $N$ even for a short suffix $u$. Computing $Q_{|u|}(u)$
or the two complete-word scores $Q_{|u|+1}(u0),Q_{|u|+1}(u1)$ does not
perform either of the sums in \eqref{eq:requiredquery}. This family of
type-uniform block assignments is not projectively consistent across
horizons: for example $Q_1(1)=1/2$, whereas the prefix marginal
$Q_4(X_1=1)=7/15$. Consistency within each fixed $Q_N$ is supplied
by summing its completions. The preceding counterexample shows why a
local-horizon substitution changes the prediction guarantee.

\subsection{Unconditional renewal generating functions}
Flajolet--Szpankowski \cite{fs02} is the closest algebraic precedent.
Their equations (7), (15), and (25) organize complete renewal words and
maximum-likelihood redundancy sums using products over gap multiplicities
and marked coefficients. Those tools are established, and truncated
coefficient multiplication itself is a standard algorithmic operation.
An unconditional normalizing coefficient has neither a supplied observed
histogram $c$ nor a constraint that the first future gap exceed an
observed age $d$. Consequently reading off that coefficient does not
evaluate \eqref{eq:requiredquery} for an arbitrary observed prefix.

The additional object here is specifically
$\prod_j(1-wz^j)^{-c_j-1}$ together with the restricted mark in
\eqref{eq:B} and the factorial multiplier in \eqref{eq:S}. These
retain the dependence between observed and future multiplicities in
$\prod_j(c_j+e_j)!/(m+k)!$. Algorithms~1--3 evaluate that object and
share it over all suffixes. This is an adaptation of classical algebra,
not a claim that generating functions or renewal types are new. The
distinction is the conditional query and its computational consequence;
no lower bound on the difficulty of adapting earlier machinery is asserted.

\subsection{Sequential CTW assignments}
CTW does supply efficiently computable sequential conditionals for its
own coding distribution. They need not equal \eqref{eq:requiredquery},
which is a different probability assignment. More generally, equality
with our $Q_N$ is unnecessary: another code with $O(\sqrt N)$ redundancy
and polynomial prefix evaluation would also solve the prediction problem.
Garivier \cite{gar06} proves worst-case CTW redundancy of order
$\sqrt N\log N$ on the renewal class. Applied directly to the same
compression-to-prediction bound, that guarantee gives only an
$O(\log n/\sqrt n)$ upper bound, rather than the desired $O(n^{-1/2})$.
Thus the missing inference is the optimal statistical guarantee, not
the existence of sequential CTW conditionals. This comparison does not
prove a statistical lower bound for every possible predictor based on CTW.

For observed finite-state Markov chains, Han--Jana--Wu \cite{hjana23}
already use suffix averages of add-one estimators, evaluated in $O(nk)$
time. The infinite-memory reduction used here is from
Han--Jiang--Wu \cite{hjw24}; the average and the reduction are not new.
The 2026 author survey \cite{hw26}, Section IV.B--C, restates the renewal
risk theorem and gives a finite-state HMM algorithm without a renewal
implementation in that discussion. The renewal example in
\cite[Appendix F]{jpr25} is an explicit limitation of its general
entropy upper bounds, not a computational lower bound.

The theorem concerns a known observation horizon, independent gaps, and
stationary initialization. Dependent gaps, noisy observations, and efficient
horizon-free updates require additional arguments. The arithmetic and bit
bounds are asymptotic computational guarantees; practical large-scale
performance is a separate question.

\appendix
\section{Independent finite verification and reproducibility}
\label{app:verification}
The accompanying standard-library Python implementation is
\texttt{anc/renewal\_predictor.py}; the single entry point
\texttt{anc/reproduce\_appendix.py} regenerates
\texttt{appendix\_results.json}. Run the latter script using Python
3.10 or later from any working directory. No training, external data,
network access, or third-party Python package is required. The
\texttt{suffix\_marginals} method returns the integer pairs $(D_t,E_t)$;
\texttt{suffix\_conditionals} and \texttt{predict\_one} return exact
\texttt{Fraction} values. The result file records the interpreter,
platform, source hashes, deterministic case names, and timing repetitions.
These are independent reference computations, not an independent human
review or a substitute for the proofs above.

\subsection{An oracle that enumerates actual type classes}
For each $N=1,\ldots,10$, the oracle enumerates all $2^N$ words and
groups them by first-one position, last-one position, and the sorted
multiset of completed gaps; the all-zero word is its own group.
It assigns every member of an enumerated group of size $r$ the weight
$N!/r$. Importantly, the reference obtains $r$ by counting actual words,
without using the factorial type-size identity, either prefix DP, or
the generating-function recurrences under test. Summing these weights
over completions produces an independent integer prefix-mass table.

The script compares every computed prefix mass with this table, including
the empty prefix and complete words. It separately checks that the
enumerated number of groups equals $M_N$ and that the empty-prefix mass
is $N!M_N$. For every nonterminal prefix it checks
\[
 S_N(u)=S_N(u0)+S_N(u1)
\]
both in the reference table and in the evaluated implementation. For
every length-$(N-1)$ history, all shared-suffix pairs are compared
individually with the reference masses; the conditional probabilities
and their average are checked exactly as rational numbers.

\subsection{Shared-array invariants and censored boundaries}
A read-only trace hook inspects the actual implementation on entry to
each nonzero-suffix query for a named boundary suite at $N=9$. A second
reference enumerates integer partitions of each span $s\le N$.
For each partition histogram $e$, it adds
$\prod_j\binom{c_j+e_j}{e_j}$ to $A_{s,k}$ and this quantity times
$\sum_{j>d}e_j$ to $B_{s,k}$. Adding one observed gap of length $d+1$
gives a separate reference for $A^+$. This reference uses direct finite
histogram enumeration and no in-place geometric multiplication or
deletion update. It compares \emph{every} retained array entry through
span $N$, including spans beyond the current query budget, as well as
$m$ and $C$. Thus leading-zero deletions, marked-gap deletions, and
growing future budgets are checked at the state level.

Named histories include all-zero and all-one words, a single one at
the beginning, middle, or end, repeated gaps of length two and three,
and repeated gaps followed by unfinished ages below, at, and above
the observed gap length. This tests the strict admissibility threshold
$j>d$, nonzero multiplicity $c_{d+1}$ when appending a one, and the
transition to all-zero suffixes. Larger named and seeded-random histories
at $N=16,32,64$ are compared with separately rebuilt prefix DPs.
These latter references do share the mathematical prefix identity;
their role is to check reuse across suffixes, not to provide a second
derivation of that identity.

\subsection{Operation growth and actual elapsed time}
The script counts visits to five kernels in the running shared-suffix
implementation: initialization, the append-one factor, denominator and
numerator query cells, and deletion. A visit entails a bounded number
of integer operations, although integer bit costs vary with $N$.
These counts are obtained from executed source lines, rather than by
returning a predicted cubic formula. On the dense all-one histories,
the script checks that their total is at most $4N^3$ for
$N=8,16,32,64$. The count excludes preprocessing, loop control, and
rational averaging, which have the separately bounded costs in
Theorem~\ref{thm:algorithm}.

For elapsed time, each run constructs a fresh predictor and computes
the complete averaged prediction with exact rational arithmetic.
Five untraced repetitions are retained, and the median is reported.
The trace-counting runs are excluded from timing. A wall-clock ratio
is machine dependent and includes increasing integer sizes and gcd
normalization in Python's \texttt{Fraction}; it need not be exactly
eight under a doubling of $N$. The asymptotic arithmetic and bit
statements follow from the proof, independently of these measurements.

\begin{table}[ht]
\centering
\caption{Executed kernel visits and untraced elapsed time. Dense all-one histories; five timing repetitions per horizon.}
\label{tab:growth}
\begin{tabular}{rrrr}
\toprule
Horizon $N$ & Kernel visits & Visits/$N^3$ & Median (ms) \\
\midrule
8 & 812 & 1.5859 & 0.217 \\
16 & 6,680 & 1.6309 & 1.573 \\
32 & 54,064 & 1.6499 & 13.488 \\
64 & 434,784 & 1.6586 & 119.323 \\
\bottomrule
\end{tabular}
\end{table}
\noindent The recorded run passed 4,082 prefix-mass comparisons, 2,036 binary mass-split checks, and 8,194 suffix-pair comparisons on 1,022 exhaustive histories. The boundary suite checked 57 complete coefficient states; the larger suite checked 1,308 additional pairs. These counts overlap across test purposes and are not independent statistical samples.
The run used Python 3.14.7 on Windows-11-10.0.26300-SP0.

\end{document}